\documentclass[aps,superscriptaddress,preprint]{revtex4}
\usepackage{amsmath,amsthm,mathtools}
\usepackage[pdftex,letterpaper=true,pagebackref=false]{hyperref}
\usepackage{hyperref,verbatim}

\usepackage{epsfig}
\usepackage{graphicx}
\usepackage{bbding,hyperref}

\usepackage{amsfonts,amssymb,bbold}
\usepackage{amsmath,amsthm}

\def\cA{\mathcal{A}}
\def\cB{\mathcal{B}}

\def\cD{\mathcal{D}}

\def\cH{\mathcal{H}}

\def\cM{\mathcal{M}}
\def\cN{\mathcal{N}}
\def\cP{\mathcal{P}}

\def\cU{\mathcal{U}}
\def\cV{\mathcal{V}}

\def\cZ{\mathcal{Z}}

\def\fB{\mathfrak{B}}

\def\fF{\mathfrak{F}}
\def\fK{\mathfrak{K}}

\def\tA{{\bf{A}}}

\def\tD{{\bf{D}}}

\def\tG{{\bf{G}}}

\def\tP{{\bf{P}}}

\def\tR{{\bf{R}}}

\def\tU{{\bf{U}}}
\def\tV{{\bf{V}}}
\def\tW{{\bf{W}}}
\def\tX{{\bf{X}}}
\def\tY{{\bf{Y}}}
\def\tZ{{\bf{Z}}}

\def\eps{\epsilon}

\def\>{\rangle}
\def\<{\langle}

\DeclareMathOperator{\tr}{Tr}

\DeclareMathOperator{\conv}{conv}
\DeclareMathOperator{\diag}{diag}

\def\tens#1{{\bf {#1}}}

\def\bbN{\mathbb{N}}

\def\bbR{\mathbb{R}}

\def\bbZ{\mathbb{Z}}
\def\bbone{\mathbb{1}}

\newcommand{\bi}[2]{\dbinom{#1}{#2}}

\newtheorem{theorem}{Theorem}[section]
\newtheorem{lemma}[theorem]{Lemma}

\newtheorem{corollary}[theorem]{Corollary}
\newtheorem{proposition}[theorem]{Proposition}

\theoremstyle{definition}

\theoremstyle{remark}
\newtheorem{remark}[theorem]{Remark}

\begin{document}
\title{Channel covariance, twirling, contraction, 
and some upper bounds on the quantum capacity}

\author{Yingkai Ouyang }
\affiliation{
\footnotesize Department of Combinatorics and Optimization, Institute of Quantum Computing, University of Waterloo, \\ 
200 University Avenue West,
\footnotesize Waterloo, Ontario N2L 3G1, Canada.\\ 
\footnotesize \texttt{y3ouyang@math.uwaterloo.ca}}
\affiliation{
\footnotesize Department of Engineering and Product Development,
Singapore University of Technology and Design,  \\ 
20 Dover Drive,
\footnotesize Singapore 138682, Singapore.\\ 
\footnotesize \texttt{yingkai\_ouyang@sutd.edu.sg}}

\begin{abstract}
Evaluating the quantum capacity of quantum channels is an important but difficult problem,
 even for channels of low input and output dimension. 
  Smith and Smolin showed that the quantum capacity of the Clifford-twirl of a qubit amplitude damping channel (a qubit depolarizing channel) has a quantum capacity that is at most the coherent information of the qubit amplitude damping channel evaluated on the maximally mixed input state.
 We restrict our attention to obtaining upper bounds on the quantum capacity using a generalization of Smith and Smolin's degradable extension technique. 
 Given a degradable channel $\cN$ and a finite projective group of unitaries $\cV$, we show that
 the $\cV$-twirl of $\cN$ has a quantum capacity at most the coherent information of $\cN$ maximized over a $\cV$-contracted space of input states. As a consequence, degradable channels that are covariant with respect to diagonal Pauli matrices have quantum capacities that are their coherent information maximized over just the diagonal input states. As an application of our main result, we supply new upper bounds on the quantum capacity of some unital and non-unital channels
 -- $d$-dimensional depolarizing channels, two-qubit locally symmetric Pauli channels, and shifted qubit depolarizing channels.
\end{abstract}

\maketitle

\section{Introduction}
The quantum capacity of a quantum channel is the maximum rate at which quantum information can be transmitted reliably across it, given arbitrarily many uses of it \cite{Wilde}. 
However, evaluating the best known regularized expressions for the quantum capacity of a general quantum channel is in general an infinite-dimensional optimization problem, and hence difficult, even for quantum channels with low dimensional input and output states. 
The quantum capacity of even the simply described family of qubit depolarizing channels is undetermined, in spite of much effort \cite{Cer00,Rai99,Rai01,SSW08,SS08,DSS97,SS07,FWh08}. 
Thus, obtaining upper bounds on the quantum capacity of quantum channels is a non-trivial and important problem.
 
Our main result generalizes the technical results of Smith and Smolin \cite{SS08} pertaining to the use of degradable extensions to obtain upper bounds on the quantum capacity of channels in terms of the coherent information of other channels. 
In our extension of Smith and Smolin's recipe, we prove that the quantum capacity of a degradable channel twirled with respect to a projective unitary group is at most the coherent information of the degradable channel maximized over a contracted input state space (Theorem \ref{thm:twirling-and-contraction}). Smith and Smolin's recipe is produced as a special case of our extension when the projective unitary group is chosen to be the full qubit Clifford group.   As a consequence, a degradable channel that is covariant with respect to diagonal Pauli matrices has a quantum capacity that is equal to its coherent information maximized over just the diagonal input states. 

As an application of our main result, we supply new upper bounds on the quantum capacity of some unital and non-unital channels
 -- $d$-dimensional depolarizing channels, two-qubit locally symmetric Pauli channels, and shifted qubit depolarizing channels.  The main ingredients that we introduce to obtain these new upper bounds are our higher dimensional amplitude damping channels that are degradable. These higher dimensional amplitude damping channels generalize qubit amplitude damping channels. 

The rest of the paper is organized in the following way. In Section \ref{sec:degradable-prelims}, we introduce notations and review concepts pertaining to the quantum capacity, degradable channels and the degradable extensions of Smith and Smolin.
In Section \ref{sec:covariance}, we review the notion of channel covariance, channel twirling, and channel contraction. In Section \ref{sec:degradable-main-result}, we present the main result of this paper, which is Theorem \ref{thm:twirling-and-contraction}, placed in the context of channel twirlings and channel covariance. 
In Section \ref{sec:degradable-applications}, we apply our main result to obtain explicit upper bounds on the quantum capacity of $d$-dimensional depolarizing channels, locally symmetric and SWAP-invariant two-qubit Pauli channels, and shifted depolarizing channels. Section \ref{sec:degradable-appendix} is our appendix, which contains the more technical ancillary results of this paper.
\section{Preliminaries}\label{sec:degradable-prelims}
\subsection{General Notation}
Given a function $f: \Omega \to \bbR$ and a subset $X \subseteq \Omega$, define the $X$-restricted convex hull of the function $f$ evaluated on the argument $x$ to be 
\[
\conv(f;x,X) := \inf_{\substack{y,z \in X \\ \lambda \in [0,1] }} \Bigl \{ 
     \lambda f(y) + (1-\lambda) f(z) : x = \lambda y + (1-\lambda)z
\Bigr \}.
\]
Given a sequence of functions $f_1 , ... ,f_n : \Omega \to \bbR$, define $ \min\{ f_1, ... , f_n \}$ to be a function that is the pointwise minimum of the sequence $f_1,... f_n$, that is,
\[
\bigl( \min\{ f_1, ... , f_n \} \bigr) (x) := \min \{f_1(x), ... , f_n(x) \}.
\]
 Now define the $X$-restricted convex hull of the sequence of functions $f_1,...,f_n$ evaluated on the argument $x$ to be
\[
\conv(f_1, ... , f_n; x,X) := \conv( \min\{ f_1, ... , f_n \}; x,X  ).
\]
We introduce this notion of the $X$-restricted convex hull because it is a tool that we later use to establish upper bounds on the quantum capacity of various quantum channels.

Define $\eta(z):= -z \log_2 z$ where $z \in [0,1]$ and $\eta(0) := 0$. Let $H_2(q) := \eta(q) + \eta(1-q)$ be the binary entropy function. Define the Pauli matrices to be
\[
  \tens{\mathbb 1} := \begin{pmatrix} 1 & 0 \\ 0 & 1 \\ \end{pmatrix},
  \tens{X} := \begin{pmatrix} 0 & 1 \\ 1 & 0 \\ \end{pmatrix},
  \tens{Z} := \begin{pmatrix} 1 & 0 \\ 0 & -1 \\ \end{pmatrix} ,
  \tens{Y} := i \tens{XZ}.
 \]
Define the Pauli group on $m$ qubits modulo phases, to be $\mathcal P_m := \{\tens{\mathbb 1}, \tens{X}, \tens{Y}, \tens{Z}\}^{\otimes m}$.  
For all $\tens P \in \mathcal P_m$, define the weight of $\tens P$ to be the number of qubits on which the
operator $\tens P$ acts non-trivially.

\subsection{Quantum Channels and the Quantum Capacity}
For a complex separable Hilbert space $\mathcal H$, let $\mathfrak B(\mathcal H)$ be the set of bounded linear operators mapping $\mathcal H$ to $\mathcal H$. In this paper, we only deal with finite-dimensional Hilbert spaces.
A quantum channel $\cN : \mathfrak B(\mathcal H_A) \to \mathfrak B(\mathcal H_B)$ is a completely positive and trace-preserving (CPT) linear map, and can be written in terms of a Kraus representation \cite{kraus}
\[
\cN(\rho) = \sum_{k} \tens{A}_k \rho \tens{A}_k^\dagger,
\]
where the completeness relation
${\sum_k \tens{A}_k^\dagger \tens{A}_k = \mathbb 1_{d_A} }$
is satisfied, ${d_A =  \dim (\mathcal H_A)}$ and $\bbone_{d_A}$ is a dimension $d_A$ identity matrix.
We can also write down the action of a quantum channel $\cN$ in terms of an isometry on the input state.
Now define an isometry $\tens W : \mathfrak B(\mathcal H_A) \to \mathfrak B(\mathcal H_E \otimes \mathcal H_B)$
\[\tens{W} = \sum_k |k\> \otimes \tens{A}_k . \]
Here $\{ |k\> \}$ is an orthonormal set, and spans a Hilbert space $\mathcal H_E$ that we interpret to be the environment. Then
\[
\tens{W} \tens{\rho} \tens{W}^\dagger = \sum_{j,k} |j\>\<k| \otimes \tens{A}_j \rho \tens{A}_k^\dagger
\]
and
\[\tr_{\mathcal H_E}(\tens{W} \tens{\rho} \tens{W}^\dagger) = \cN(\tens{\rho}).\]
Then we can define the {\bf complementary channel} $\cN^C : \mathfrak B (\mathcal H_A) \to \mathfrak B( \mathcal H_E)$ \cite{DeS03} as
\[
\cN^C(\tens{\rho}) = \tr_{\mathcal H_B}(\tens{W}\rho \tens{W^\dagger}).
\]
Since we are free to choose the orthonormal basis of the environment $\mathcal H_E$, $\cN^C$  is only unique up to a unitary transformation. We use the above definition as our canonical one.
Let $\cN^C(\tens{\rho}) =  \sum_\mu \tens{R_\mu } \tens{\rho } \tens{R_\mu}^\dagger$.
The $j$-th row of $\tens{R_{\mu}}$ is the $\mu$-th row of $\tens{A}_j$, where $\tR_\mu = \sum_j |j\>\<\mu| \tA_j $ \cite{KMNR06}. To see this, observe that
\begin{align*}
\cN^C(\rho)
&= \tr_{\mathcal H_B}(\tens{W} \tens{\rho} \tens{W}^\dagger)\notag\\
&= \tr_{\mathcal H_B}
\Bigl(\sum_{j,k} |j\>\<k| \otimes \tens{A}_j \rho \tens{A}_k^\dagger
\Bigr)\notag\\
&= \sum_{j,k} |j\>\<k| \tr
\left( \tens{A}_j \rho \tens{A}_k^\dagger
\right)\notag\\
&= \sum_{j,k} |j\> \sum_{\mu}\<\mu|
\left( \tens{A}_j \rho \tens{A}_k^\dagger
\right) |\mu\> \<k|\notag\\
&= \sum_{\mu}   \Bigl(\sum_j |j\>\<\mu| \tens{A}_j \Bigr)   \rho \Bigl(\sum_k \tens{A}_k^\dagger |\mu\> \<k|\Bigr)\notag\\
&= \sum_{\mu}  \tens{R_\mu}   \rho \tR_\mu^\dagger.
\end{align*}
For a quantum channel $\cN : \mathfrak B (\mathcal H_A) \to  \mathfrak B(\mathcal H_B)$, Schumacher and Nielsen defined its coherent information \cite{ScN96} with respect to an input state as a difference of von Neumann entropies
\[I_{\rm coh}(\cN, \rho) :=  S(\cN(\rho)) - S(\cN^C(\rho)) \]
where the von Neumann entropy of a state $\rho$ is 
\[
S(\rho) := -\tr (\rho \log_2 \rho).
\]
We denote the channel's optimized coherent information as
\[I_{\rm coh}(\cN) := \max_{\rho}  I_{\rm coh}(\cN, \rho). \]
Here, the maximization of $\rho$ is performed over all quantum states in $\fB(\cH_A)$.
Lloyd \cite{Llo97}, Shor \cite{Sho02} and Devetak \cite{Dev03} showed that the quantum capacity of $\cN$ is
\begin{align}
Q(\cN) =   \lim_{n\to \infty} \frac{1}{n} I_{\rm coh}(\cN^{\otimes n})  ,
\label{eq:quantum-capacity}
\end{align}
and the limit on the right hand side of (\ref{eq:quantum-capacity}) exists \cite{BNS97}. Schumacher and Westmoreland also demonstrated that a channel's coherent information is also a lower bound on the amount of its private information \cite{SWe98}.
\subsection{Degradable Channels and Degradable Extensions}
A channel $\mathcal N$ is {\em degradable} \cite{DeS03} if it can be composed with another quantum channel $\Psi$ to become equivalent to its complementary channel $\mathcal N^C$, that is $\cN^C = \Psi \circ \cN$. 
Physically, this means that the environment associated with the channel $\cN$ can be simulated using the output quantum state of channel the $\cN$. 
Conversely, $\cN$ is {\em antidegradable} if its complementary channel $\mathcal N^C$ is degradable. A channel $\mathcal N_{\rm ext}$ is a {\em degradable extension} \cite{SS08} of channel $\mathcal N$ if $\mathcal N_{\rm ext}$ is degradable and there exists a quantum operation $\Psi$ such that $\Psi \circ \mathcal N_{\rm ext} = \mathcal N$.

A degradable channel $\cN$ has a simple expression for its quantum capacity, which is $Q(\cN) = I_{\rm coh}(\cN)$ \cite{DeS03}. If the degradable channel $\cN$ also extends a channel $\cM$ that is not necessarily degradable, we have $Q(\cM) \le I_{\rm coh}(\cN)$. Moreover if $\cN = \sum_i \lambda_i \cN_i$ is a convex combination of degradable channels $\cN_i$, then we have the crucial convexity property \cite{SS08} given by
\[
Q(\cM) \le \sum_{i} \lambda_i  I_{\rm coh}(\cN_i).
\]
Thus, degradable extensions can be used to construct upper bounds on the quantum capacity of quantum channels \cite{SS08}. 
\section{Channel Covariance, Twirling and Contraction} 
\label{sec:covariance}
In this section, we introduce the concepts of covariance, twirling and contraction which are essential to state our main result in Theorem \ref{thm:twirling-and-contraction}.

Let $\cV$ be a set of unitary operators 
A channel $\cN$ is said to be {\em $\cV$-covariant} if for all input quantum states $\rho$ and elements $\tV$ of $\cV$, we have
$
    \cN(\tV \tens{\rho} \tV^\dagger) = \tV \cN(\tens{\rho}) \tV^\dagger.
$
 Properties of quantum channels covariant with respect to locally compact groups were studied by Holevo \cite{Hol93}. 
 
Define
$
\displaystyle
\cV_\rhd (\rho) := \frac{1}{|\cV|} \sum_{\tens V \in \cV} \tens V \rho \tens V^\dagger
$
to be a {\em $\cV$-contraction channel}. 
We also denote the {\em $\cV$-twirl} of $\cN$ as the channel $\cN_{\ltimes \cV \rtimes}$ where
$ \displaystyle
\cN_{\ltimes \cV \rtimes}(\rho) := 
\frac{1}{|\cV| } \sum_{\tV \in \cV}  \tV^\dagger
 \cN(\tV \tens{\rho} \tV^\dagger )
 \tV.
$
When the set $\cV$ is the $m$-qubit Pauli set $\cP_m$, the $\cV$-twirl of a channel $\cN$ has the Kraus operators
\[
\frac{\tens P}{2^m} \sqrt{  \sum_{\tens K \in \fK_\cN } 
 \Bigl|\tr(\tens{P K})  \Bigr|^2     }  \label{eq:paulidiag},
 \]
where $\tP \in \cP_m$ and $\fK_\cN$ is the Kraus set of $\cN$ \cite{CCEE09}. 

We say that a finite set of unitary matrices $\cV$ is a finite projective group if (i) no two distinct elements of $\cV$ are equivalent up to a constant, and (ii) for all $\tV$ and $\tW$ in the set $\cV$, there exists a unique complex number of unit magnitude $z_{\tV, \tW^\dagger}$ such that 
$z_{\tV, \tW^\dagger }  \tV \tW^\dagger$ is also an element of $\cV$. A channel that is $\cV$-covariant need not be invariant under $\cV$-twirling. However this is the case when $\cV$ is a multiplicative (or projective) group $\cV$.

\section{Main Result}  \label{sec:degradable-main-result}
The main result of this paper is a generalization of Smith and Smolin's technique of degradable extensions (see Lemma 8 of \cite{SS08}). Our main result states that the quantum capacity of a $\cV$-twirled degradable channel is at most its coherent information maximized over the set of correspondingly $\cV$-contracted input states. Here $\cV$ is a finite projective group of $d$-dimensional unitary operators. Our result is a generalization of Smith and Smolin's technique in the sense that the set $\cV$ need not be restricted to just the set of single-qubit Clifford operators.

To state our main result formally, first define $\widetilde \cN$ to be an extension of the $\cV$-twirl $\cN_{\ltimes \cV \rtimes}$, where
 \begin{align}
\widetilde{\cN} (\rho) 
       &:= 
              \sum_{\tens V \in \cV} \frac{1}{|\cV|}  
                   \tV^\dagger \cN (\tV \rho \tV^\dagger ) \tV 
                         \otimes 
                  |\tens V\>\<\tens V| \label{eq:tildeN}.
\end{align}
\begin{theorem}[Twirling and Contraction]
 \label{thm:twirling-and-contraction}
 Let $\cV$ be a projective group of $d$-dimensional unitary matrices, $\mathcal N$ be a degradable channel with $d$-dimensional input and output states, and $\widetilde \cN$ be as defined in (\ref{eq:tildeN}). Then 
$
Q(\cN_{\ltimes \cV \rtimes}) \le Q(\widetilde \cN) \le \max_{\rho}I_{\rm coh}(\cN, \cV_\rhd(\rho) ).
$
\end{theorem}
We supply the proof of Theorem \ref{thm:twirling-and-contraction} in Section \ref{sec:proof-thm:twirling-and-contraction}. The main idea of the proof is a straightforward extension of the methods used by Smith and Smolin (Lemma 8 in \cite{SS08}). A technical result needed in the proof is the following proposition.
\begin{proposition}\label{prop:twirled-extension-C}
Let $\cN$ be a quantum channel with $d$-dimensional input and output states, $\cV$ be a set of $d$-dimensional unitary matrices, and $\widetilde \cN$ be as defined in (\ref{eq:tildeN}). Then 
\begin{align}
\widetilde \cN^C(\rho) = 
\sum_{\tV \in \cV } \frac{1}{|\cV|}
     \cN^C (\tV \rho \tV^\dagger)  \otimes |\tV\>\<\tV| . \label{eq:tildeN-C}
\end{align}
\end{proposition}
The proof of Proposition \ref{prop:twirled-extension-C} uses only techniques from \cite{SS08}, and we defer its proof to Section \ref{sec:proof-prop-twirled-ext}. 
\begin{corollary}[Degradable and Covariant Channels]\label{coro:degradable-and-covariant}
Let $\cV$ be a finite projective unitary group. If a degradable channel $\cN$ is also $\cV$-covariant, then 
$
Q(\cN) = \max_\rho I_{\rm coh}(\cN, \cV_\rhd(\rho) ).
$
\end{corollary}
\begin{proof}[Proof of Corollary \ref{coro:degradable-and-covariant}]
Since the channel $\cN$ is degradable, 
\[Q(\cN) = I_{\rm coh}(\cN) \ge \max_\rho I_{\rm coh}(\cN, \cV_\rhd(\rho) ).\] Since 
${\cN_{\ltimes \cV \rtimes} = \cN}$, Theorem \ref{thm:twirling-and-contraction} implies that 
$Q(\cN) \le \max_\rho I_{\rm coh}(\cN, \cV_\rhd(\rho) ).$ 
\end{proof}
The set of diagonal $m$-qubit Pauli matrices $\cZ_m := \{ \bbone, \tZ \}^{\otimes m}$ is an example of a finite projective group of unitary matrices.
Our result shows that a degradable channel $\cN$ which is  $\cZ_m$-covariant has quantum capacity equal to $I_{\rm coh}(\cN, \rho)$ maximized over all diagonal $m$-qubit quantum states $\rho$. 

\subsection{Examples of Degradable Channels that are Covariant} \label{subsec:special}
Here, we show that examples of degradable channels that are $\cZ_m$-covariant include special $m$-qubit Hadamard channels that admit a Pauli decomposition with diagonal Kraus operators, all $m$-qubit almost-Pauli channels, all single-qubit degradable channels, and the higher dimensional amplitude damping channels that we introduce in Section \ref{sec:new-AD-channels}. We prove these facts in this section.

We say that a quantum channel is {\em almost-Pauli} if it admits a Kraus decomposition with all of its Kraus operators having the form
$\tens{K}_j = \tD_j \tens{P }_j$
where $\tD_j$ is a size $2^m$ diagonal matrix and $\tens{P}_j \in \mathcal P_m$. 
{Almost-Pauli} channels are covariant with respect to the $m$-qubit diagonal Pauli matrices because
\[
(\tD_j \tP_j)( \Lambda \tW \Lambda) ( \tP_j \tD_j^\dagger) =
\Lambda(\tD_j \tP_j) \tW  ( \tP_j \tD_j^\dagger)\Lambda
\]
 for all Paulis $\tW$ and diagonal Paulis $\Lambda \in \{ \bbone, \tZ\}^{\otimes m}$. The above equality holds because we can `propagate' the $\Lambda$'s `outwards'. This is because Pauli matrices either commute or anti-commute under multiplication, and diagonal matrices commute under multiplication. Hence a degradable almost-Pauli channel is $\cZ_m$-covariant.
\begin{proposition}
Qubit degradable channels are $\cZ_1$-covariant.
\end{proposition}
\begin{proof}
All qubit degradable channels necessarily have Kraus operators of the following form \cite{WPG08, CRS08}
\[
\begin{pmatrix} \cos \alpha & 0 \\ 0 & \cos \beta \\ \end{pmatrix}, \quad
\begin{pmatrix} 0 & \sin \beta \\ \sin \alpha & 0 \\ \end{pmatrix} =  \begin{pmatrix} \sin \beta & 0 \\ 0 & \sin \alpha  \\ \end{pmatrix} \tens{X}.
\]
Hence these channels are almost-Pauli and the result follows. 
\end{proof}

Any Hadamard channel maps a quantum state to some Hadamard product of it, and is the complementary channel of an entanglement breaking channel (which admits a Kraus decomposition with Kraus operators of rank one) (see \cite{KMNR06,BHTW10} and the references therein). Consider a special almost-Pauli $m$-qubit channel $\cA$ with only diagonal Kraus operators 
${\tA_i = \sum_{j \in \bbZ_{2^m}} c_{i,j}|j\>\<j|}$
 for ${i \in \bbZ_{2^m}}$. 
 This channel $\cA$ is Hadamard because its complementary channel $\cA^C$ has Kraus operators 
 ${R_{\mu} = \sum_{i \in \bbZ_{2^m}} |i\>\<\mu| c_{i, \mu}}$ of column rank at most one.
Hence almost-Pauli channels with purely diagonal Kraus operators are examples of Hadamard channels that are also {$\cZ_m$-covariant}.

\subsection{Proof of Theorem \ref{thm:twirling-and-contraction}}
\label{sec:proof-thm:twirling-and-contraction}
Since $\cV$ is a finite projective unitary group, for all $\tV, \tW \in \cV$, there exists a unique phase constant $z_{\tV, \tW} \in \bbR$ such that $z_{\tV, \tW} \in \cV$. Define the $\star$ product to be the binary operation given by 
\[
\tV \star \tW := z_{\tV, \tW} \tV \tW.
\]
Hence for all $\tV, \tW \in \cV,$ we also have $\tV \star \tW \in \cV$. Since every element of $\cV$ is a unitary matrix and $\cV$ is also a group, $\tV \in \cV$ also implies that $\tV^{\dagger} \in \cV$.  Hence if ${\tR = \tV \star \tW}$, we also have ${\tV = \tR \star \tW^\dagger}$.
 \begin{align}
&\widetilde{\cN} (\tW \rho \tW^\dagger)= 
\sum_{\tV \in \cV} \frac{1}{|\cV|} 
(\tW \tW^\dagger) \tV^\dagger 
\cN(\tV 
	\tW \rho \tW^\dagger 
\tV^\dagger) \tV
( \tW \tW^\dagger) \otimes |\tV\>\< \tV|
  \notag \\
&= \sum_{\tV \in \cV} \frac{1}{|\cV|} 
\tW
(z_{\tV,\tW} \tV \tW)^\dagger 
\cN((z_{\tV,\tW}  \tV \tW) 
\rho 
(z_{\tV,\tW}  \tV \tW)^\dagger )
(z_{\tV, \tW} \tV \tW) \tW^\dagger
\otimes |\tV\>\< \tV| \notag .
 \end{align}
 Making the substitution $\tR = \tV \star \tW = z_{\tV, \tW} \tV \tW$ we get
 \begin{align}
\widetilde{\cN} (\tW \rho \tW^\dagger) =& 
 \sum_{\tR \star \tW^\dagger \in \cV} \frac{1}{|\cV|} 
 \tW \tens R^\dagger
 \cN(\tens R \rho \tens R^\dagger ) 
 \tens R \tW^\dagger
  \otimes |\tR \star \tW^\dagger \>\< \tR \star \tW^\dagger| \notag.
  \end{align}
Let $\tens U_{\tW} := \sum_{\tR \in \cV} |\tR \star \tW^\dagger \>\<\tR |$ be a unitary matrix that depends on $\tW \in \cV$. Since $\cV$ is a group under the binary operation $\star$, $\cV \star \tW = \cV$, and we can replace the summation index of the right hand side of the above equation to get
  \begin{align}
  \widetilde{\cN} (\tW \rho \tW^\dagger) =& 
 (\tW \otimes \tU_{\tW}) 
		\widetilde{\mathcal N} (\rho )
	(\tW^\dagger \otimes \tU_{\tW}^\dagger) \label{eq:lem8}.
 \end{align}

Now we can use the isometric extensions of the channels $\cN$ and $\widetilde \cN$ to show that (see Proposition \ref{prop:twirled-extension-C})
\begin{align*}
\widetilde {\cN}^C(\rho)
&= \sum_{\tens V \in \cV} \frac{1}{|\cV|}    \cN^C(\tV \rho \tV^\dagger ) \otimes |\tens V\>\<\tens V|.
\end{align*}
By a similar argument as in (\ref{eq:lem8}),
 \begin{align}
   \widetilde \cN^C (\tens V\rho \tens V^\dagger) =
   (\bbone_{d_E} \otimes \tens U_{\tens V}) \widetilde \cN^C (\rho )(\bbone_{d_E} \otimes \tens U_{\tens V}^\dagger), \label{eq:ncinv}
 \end{align}
 where $d_E$ is the dimension of the output states of the complementary channel $\cN^C$. Note that the von Neumann entropy is additive with respect to each block in a block diagonal matrix, and is also invariant under unitary conjugation of its argument. Hence the coherent information of the degradable extension $\widetilde \cN$ evaluated on the input state $\rho$ is
 \begin{align*}
 S(\widetilde \cN(\rho) ) - S(\widetilde \cN^C(\rho)) 
 &=\left( \sum_{\tV \in \cV} \frac{1}{|\cV|} 
          S\bigl(\cN(\tV \rho \tV^\dagger) \bigr)
    \right)
        -
    \left( \sum_{\tV \in \cV} \frac{1}{|\cV|} 
          S\bigl(\cN^C(\tV \rho \tV^\dagger) \bigr)
    \right)\notag\\
    &= 
      \sum_{\tV \in \cV} \frac{1}{|\cV|} 
          I_{\rm coh} \bigl(\cN ,   \tV \rho \tV^\dagger \bigr) \notag\\
    &\le I_{\rm coh} \Bigl(\cN ,
      \sum_{\tV \in \cV} \frac{1}{|\cV|} 
              \tV \rho \tV^\dagger \Bigr) ,
 \end{align*}
where the inequality above results from the concavity of the coherent information of degradable channels with respect to the input state \cite{YHD08}. Hence the coherent information of the degradable channel $\cN$ maximized over all output states of the $\cV$-contraction channel upper bounds the coherent information and the quantum capacity of the degradable extension $\widetilde \cN$. \hfill $\qed$

\section{Application to obtain Upper Bounds} \label{sec:degradable-applications}
\subsection{Degradable Amplitude Damping Channels} \label{sec:new-AD-channels}
Qubit amplitude damping channels model spontaneous decay in two-level quantum systems \cite{nielsen-chuang}, and also model the map induced by beamsplitter acting on a superposition of the vacuum and a single photon with a trace taken over one output mode \cite{GFa05}. Hence knowledge of their quantum capacity is a physically relevant problem. 
These channels (when degradable) are essential ingredients of Smith and Smolin's recipe \cite{SS08} for upper bounding the quantum capacity of the qubit depolarizing channel \cite{SS08}. Analogously, higher dimensional generalizations of the qubit amplitude damping channel that are degradable are essential ingredients of Theorem \ref{thm:twirling-and-contraction} in upper bounding the quantum capacity of higher dimensional channels.

In this section, we introduce {uniform amplitude damping channels} and special two-qubit amplitude damping channels which generalize the single-qubit amplitude damping channels. We also introduce the beamsplitter-type amplitude damping channel that models multi-photon input states passing through a beamsplitter. We give sufficient conditions for these channels to be degradable.

Define a {\em uniform amplitude damping channel} $\cA_{\gamma ,d}$ to be a channel with the Kraus operators 
$
|0\>\<0| + \sum_{i=1}^{d-1}\sqrt{1-\gamma}|i\>\<i|$ and $
\sqrt{\gamma} |0\>\< j |$, where $1\le j \le d-1.$
\begin{proposition}\label{lem:UAD-degradable}
Let integer $d\ge 2$, and $0 \le \gamma \le \frac{1}{2}$. Then $\cA_{\gamma,d}$ is a degradable channel.
\end{proposition}
\begin{proof}
Note that $\cA_{\frac{1-2\gamma}{1-\gamma}, d} \circ \cA_{\gamma,d} = \cA_{1-\gamma, d} = \cA_{\gamma,d}^C$.
\end{proof}
We assume that a beamsplitter of transmissivity $\eta \in [0,1]$ is a unitary operation $\cU_{\rm BS, \eta}$ that (1) maps an input bosonic mode (with annihilation operator $a$ and Hilbert space $A$) and an external bosonic mode (with annihilation operator $b$ and Hilbert space $B$) to two output bosonic modes \cite{GFa05,GerryKnight},
(2) maps a tensor product of vacuum states to a tensor product of vacuum states \cite{GerryKnight}, 
and (3) maps the annihilation operators $a$ and $b$ to 
$ \eta a+ \sqrt{1-\eta} b$  and 
$ \sqrt{1-\eta } a + \eta b$  respectively. We also assume that the input state in the external bosonic mode is the vacuum state $(|0\>\<0|)_B$.
If these assumptions are self-consistent, one can show that 
$\tr_B( \cU_{{\rm BS}, 1-\gamma}( (|i\>\<j|)_A \otimes (|0\>\<0|)_B ))$ is equivalent to the expression on the right hand side of (\ref{eq:BS-AD}) summed over all natural numbers $n$ for all $i,j \in \bbN$. This motives us to define a {\em beamsplitter-type amplitude damping channel} $\cA_{{\rm BS},\gamma}$ to be a channel with Kraus operators 
\[A_{n, \gamma} := \sum_{k \ge 0} |k\>\<k+n| 
\sqrt{\bi{n+k}{n} \gamma^n (1-\gamma)^k } \]
for all $n \in \bbN$ and $\gamma \in [0,1]$. 
\begin{proposition}\label{prop:BS-AD-degradable}
Let $0 \le \gamma \le \frac{1}{2}$. Then $\cA_{{\rm BS}, \gamma}$ is a degradable channel.
\end{proposition}
\begin{proof}
Note that for all $i,j,n \in \bbN$ and $0 \le \gamma < 1$,
\begin{align}
A_{n,\gamma}|i\>\<j| A_{n,\gamma} ^\dagger
= 
|i-n\>\<j-n| 
	\left(\frac{\gamma}{1-\gamma} \right)^{n}
	\sqrt{\bi{i}{n} \bi{j}{n} 
	(1-\gamma)^{i+j}
}.\label{eq:BS-AD}
\end{align}•
Let $\gamma_1, \gamma_2 \in [0,\frac{1}{2}]$. The identity 
$\bi{i}{n_1}\bi{j}{n_1} \bi{i-n_1}{n_2}\bi{j-n_1}{n_2} = 
\bi{n}{n_1}^2 \bi{i}{n} \bi{j}{n}$ and the above equation implies that 
\begin{align}
&\sum_{\substack{n_1 + n_2 = n \\ n_1, n_2 \in \bbN} }
	A_{n_2,\gamma_2} A_{n_1,\gamma_1} 	
		|i\>\<j|
	A_{n_1,\gamma_1}^\dagger A_{n_2,\gamma_2}^\dagger \notag\\
= &\sum_{n_1=0}^n \bi{n}{n_1} 
	\left(\frac{\gamma_1}{1-\gamma_1} \right)^{n_1}
	\left(\frac{\gamma_2}{1-\gamma_2} \right)^{n-n_1}
	\sqrt{\bi{i}{n} \bi{j}{n} 
		(1-\gamma_1)^{i+j}
		(1-\gamma_2)^{i+j}
	}\notag\\
=& A_{n,\gamma_1 + \gamma_2 - \gamma_1 \gamma_2}
	|i\>\<j|
 A_{n,\gamma_1 + \gamma_2 - \gamma_1 \gamma_2}^\dagger, \notag
\end{align}
from which it follows that 
$\cA_{{\rm BS},\frac{1-2\gamma}{1-\gamma}}
 \circ 
\cA_{{\rm BS},\gamma} = \cA_{{\rm BS},1-\gamma} = \cA_{{\rm BS},\gamma}^C$.
\end{proof}
Let $s_1 = \sqrt{1- x}$ and $s_2 = \sqrt{1-2y-z}$. For $x,y ,z \ge 0$ and $1-2y-z \ge 0$, we define $\cA_{x,y,z} $ to be a channel with the Kraus operators 
\begin{align}
\tens{A}_0 &= |0\>\<0| +s_1 (|1\>\<1| + |2\>\<2|) + s_2 |3\>\<3| \notag\\
\tens{A}_1 &= \sqrt{x} |0\>\<1| + \sqrt{y} |2\>\<3|\notag\\
\tens{A}_2 &= \sqrt{x} |0\>\<2| + \sqrt{y} |1\>\<3| \notag\\
\tens{A}_3 &= \sqrt{z} |0\>\<3| \label{channel-definition}.
\end{align}
Observe that $\cA_{z,0,z} = \cA_{z,4}$, and hence the channels $\cA_{x,y,z}$ generalize the uniform amplitude damping channels of dimension four.

Define the set
 \begin{align}
 \fF_{x,y,z} = \Bigl\{
(x,y,z) &\ge 0:
\quad 2y+z<1, 
\quad x < \frac{1}{2}, \quad
2z \le1-2y\Bigl(2-\frac{x}{1-x}\Bigr) \Bigr\}.
\label{ineq:Phixyz}
\end{align}
\begin{lemma}\label{lem:Phixyzdegradable}
Let $(x,y,z) \in \fF_{x,y,z}$. Then $\cA_{x,y,z}$ is a degradable channel with degrading map $\cA_{g,h,k}$, where
\begin{align}
g &= \frac{1-2x}{1-x}, \quad h = \frac{gy}{(1-2y-z)}\notag\\
k &= 1-2h - \frac{z}{1-2y-z}  \label{eq:ghk} .
\end{align}
\end{lemma}
We supply the proof of Lemma \ref{lem:Phixyzdegradable} in Section \ref{sec:proof-degradable}.
Our special two-qubit channels are also $\cZ_2$-covariant, which simplifies the evaluation of their quantum capacities via use of Corollary \ref{coro:degradable-and-covariant} when they are also degradable.
\begin{proposition} \label{prop:specialchannel}
If the linear map $\cA_{x,y,z}$ defined by (\ref{channel-definition}) is a quantum channel, then it is also $\cZ_2$-covariant.
\end{proposition}
\begin{proof}
It suffices to show that  $\cA_{x,y,z}$ is almost-Pauli. This means that we have to show that every Kraus operator of $\cA_{x,y,z}$ can be written in the form
$\tens{K}_i = \tD_i \tens{P}_i $
where $\tD_i$ is diagonal and $\tens{P}_i$ is a two-qubit Pauli. We define the vectors $|0\>,|1\>, |2\>, |3\>$ to be the two qubit states $|0,0\> , |0,1\>, |1,0\> , |1,1\>$ respectively. One can verify using equations (\ref{n01}), (\ref{n23}), (\ref{n02}), (\ref{n13}), (\ref{n03}) that a suitable choice of the matrices $\tD_i$ and $ \tens{P}_i$ is given by
\begin{align}
\tD_0 &= \sum_{i=0}^3 a_{0,i} |i\>\<i| ,\quad  &\tens{P}_0 = \tens{\mathbb 1} \otimes \tens{\mathbb 1} \notag\\
\tD_1 &= a_{1,1}|0\>\<0| - a_{1,2} |2\>\<2|, \quad &\tens{P}_1 = \tens{Z} \otimes \tens{X} \notag\\
\tD_2 &= a_{2,1}|0\>\<0| - a_{2,2}|1\>\<1| , \quad &\tens{P}_2 = \tens{X} \otimes \tens{Z} \notag\\
\tD_3 &= |0\>\<0| , \quad &\tens{P_3} = \tens{X} \otimes \tens{X}. \notag  & 
\end{align}
\end{proof}

\subsection{$d$-dimensional Depolarizing Channels}
The $d$-dimensional depolarizing channel of depolarizing probability $p$ can be described as a quantum channel that maps an $d$-dimensional input state to a convex combination of the maximally mixed $d$-dimensional state and the input state, and is defined as
 \[
 \mathcal D_{p,d}(\tens{\rho}) = \tens{\rho} \left( 1- p \frac{d^2-1}{d^2} \right) + \frac{\tens{\mathbb 1}_d}{d} \left(p \frac{d^2-1}{d^2}\right) \tr(\rho).
 \]
 Upper bounds \cite{Cer00,Rai99,Rai01,SSW08,SS08} and lower bounds \cite{Ace97,DSS97,SS07,FWh08} on the quantum capacity of qubit depolarizing channels, the simplest type of depolarizing channels, have been studied. However these bounds are not tight when the depolarizing probability is in the interval $(0,\frac{1}{4})$. 
Even less is known about the quantum capacity of higher dimensional depolarizing channels.
The goal of this section is to tighten the upper bounds for the quantum capacity of $d$-dimensional depolarizing channels.

The obvious upper bounds for the quantum capacity of depolarizing channels come from combining Cerf's no-cloning bounds \cite{Cer00} with Smith and Smolin's technique \cite{SS08}. By Cerf's result, a $d$-dimensional depolarizing channel of depolarizing probability $p$ is both degradable and anti-degradable when
\begin{align}
p=\frac{d}{2d+2}\frac{d^2-1}{d^2}= \frac{d^2-1}{2d(d+1)} = \frac{d-1}{2d}. \label{eq:nocloning}
\end{align}
Hence applying Smith and Smolin's technique of degradable extensions \cite{SS08} immediately gives the upper bound of 
\begin{align}
Q(\cD_{p,d}) \le (\log_2 d)
\left(1- p \frac{2d}{d-1} \right)
\end{align}
for depolarizing probability $0 \le p \le \frac{2d}{d-1}$. We call this upper bound the no-cloning upper bound for the quantum capacity of the depolarizing channel.

Conversely, an obvious lower bound for the quantum capacity of the $d$-dimensional depolarizing channel of noise strength $p$ is $\max(0,\log_2 d +(1-p)\log_2(1-p) + p \log_2(\frac{p}{d^2-1}))$, 
which is the maximum of zero and its coherent information evaluated on the maximally mixed state. 

\begin{figure}[h!]
  \centering
    \includegraphics[width=1.0\textwidth]{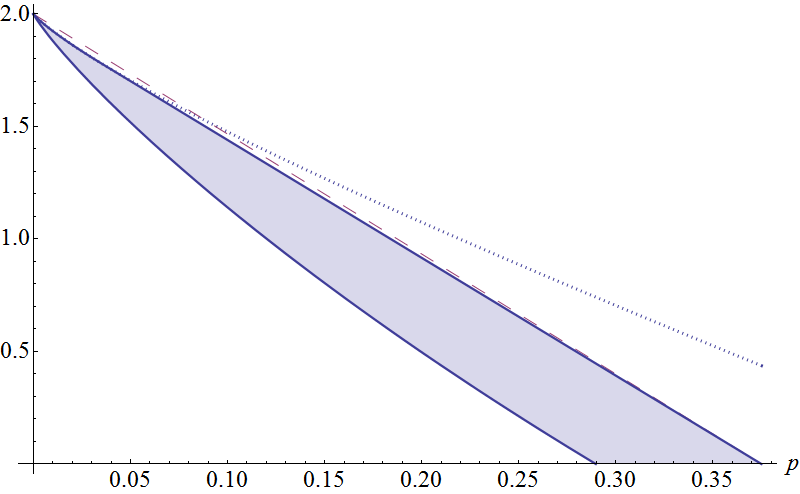}
      \caption{The upper and lower boundaries of the shaded region depict the upper and lower bounds for $Q(\cD_{p,4})$. The dotted line and dashed lines are upper bounds that comes from Cerf's no-cloning bound and our uniform amplitude damping channel  respectively (see Theorem \ref{thm:depol4}).
      }\label{fig:depol}
\end{figure}

The following theorem gives our upper bound on the quantum capacity of 
$d$-dimensional depolarizing channels. We depict our upper bound for the two-qubit case in Figure \ref{fig:depol}.
\begin{theorem} \label{thm:depol4}
For integers $d$ at least two and $0 \le p \le \frac{d-1}{2d}$,
\[
Q(\cD_{p,d}) \le
\conv\Bigl( f_1,f_2; p,  [0 , \frac{d-1}{2d}] \Bigr)
\]
where 
\[f_1(p) = I_{\rm coh} \Bigl(
\cA_{\frac{2d}{(d-1)^2}  (\sqrt{1-p}-(1-\frac{pd}{2}))  ,d},   \frac{\bbone}{d} \Bigr) \]
 and 
\[
f_2(p) = 
 \Bigl( 
      1 - p\frac{2d}{d-1}
 \Bigr)
\log_2 d.
 \]
\end{theorem}
\begin{remark}
To evaluate the upper bound of the theorem above, note that
\begin{align*}
 I_{\rm coh} \Bigl(\cA_{\gamma,d}, \frac{\bbone}{d} \Bigr) = 
 \eta\Bigl( \frac{1+(d-1)\gamma}{d} \Bigr) + (d-1)\eta\Bigl( \frac{1-\gamma}{d} \Bigr) -\eta\Bigl(1- \frac{(d-1) \gamma}{d} \Bigr) 
 - (d-1) \eta\Bigl( \frac{ \gamma}{d} \Bigr).
\end{align*}
\end{remark}
\begin{proof}[Proof of Theorem \ref{thm:depol4}]
The channel $\cA_{\gamma,d}$ has exactly one Kraus operator of non-zero trace equal to $1+(d-1)\sqrt{1-\gamma}$. Hence the complete Clifford-twirl of $\cA_{\gamma,d}$ is $\cD_{p,d}$, where
$
1-p = 
\left(\frac{1+(d-1) \sqrt{1-\gamma}}{d}\right)^2. 
$
The non-negative solution for $\gamma$ of the preceding equation for feasible values of $p$ and $d$ gives
$
\gamma = \frac{2d}{(d-1)^2}\Bigl(\sqrt{1-p}-(1-\frac{pd}{2})\Bigr)
$
as required. Hence with Theorem \ref{thm:twirling-and-contraction}, we have the bound $Q(\cD_{p,d}) \le I_{\rm coh}(\cA_{\gamma,d}, \frac{\bbone}{4})$. Cerf's no-cloning bound also gives $Q(\cD_{p,d}) \le f_2(p)$. The convexity of upper bounds obtained from degradable extensions then gives the result.
\hfill 
\end{proof}
\subsection{Two-Qubit Pauli Channels}
The tensor product of a pair of qubit Pauli channels is a two-qubit Pauli channel, but conversely a two-qubit Pauli channel need not admit a tensor product decomposition into a pair of qubit Pauli channels. The two-qubit Pauli channels that we study are invariant under the SWAP operation, and local Clifford twirling. We call such channels $(q_1,q_2)$-channels; these channels apply weight $i$ Paulis from $\cP_2$ with probabilities $q_i$.

To obtain upper bounds on the quantum capacity of $(q_1,q_2)$-channels, we first consider the equalities
\begin{align}
q_1 &=
\frac{(1-\sqrt{1-2y-z})^2}{8} + \frac{(\sqrt{x} + \sqrt{y})^2}{4}
\notag\\
q_2 &=
\frac{(1-2\sqrt{1-x} + \sqrt{1-2y-z})^2}{16} + \frac{(\sqrt{x} - \sqrt{y})^2}{4}
+ \frac{z}{4}
. \label{eq:q1q2-xyz}
\end{align}
\begin{theorem} \label{thm:two-qubit-pauli}
Let $q_1 \in [0,0.2]$ and $q_2 \in [0,0.3]$.
Then the quantum capacity of a $(q_1,q_2)$-channel is at most 
$
\conv \Bigl( f ; 　(q_1,q_2), [0, 0.2] \times [0, 0.3] \Bigr) 
$ 
where $f((q_1,q_2))$ is the infimum of 
$ I_{\rm coh}(\cA_{x,y,z}, \frac{\bbone}{4})$ over the vectors $(x,y,z)$ in $\fF_{x,y,z}$ that satisfy (\ref{eq:q1q2-xyz}).
\end{theorem}
\begin{remark}
To evaluate the upper bound in the theorem above, note that
\begin{align}
I_{\rm coh}(\cA_{x,y,z}, \tfrac{\bbone}{4}) = 
&  \eta\Bigl( \frac{1+2x+z}{4} \Bigr)
 + 2\eta\Bigl( \frac{1-x+y}{4}  \Bigr)
  + \eta\Bigl( \frac{1-2y-z}{4}  \Bigr)\notag\\
& \quad
 -\eta\Bigl(1- \frac{2x+2y+z}{4}  \Bigr)
  - 2\eta\Bigl( \frac{x+y}{4}  \Bigr)
   -  \eta\Bigl( \frac{z}{4}  \Bigr).
\end{align}
\end{remark}
\begin{proof}[Proof of Theorem \ref{thm:two-qubit-pauli}]
Let $(x,y,z)$ be a vector in $\fF_{x,y,z}$ that satisfies (\ref{eq:q1q2-xyz}). Then $\cA_{x,y,z}$ is a degradable channel (Lemma \ref{lem:Phixyzdegradable}), and can be twirled to become a $(q_1,q_2)$-channel (Proposition \ref{prop:twirling}). The use of Theorem \ref{thm:twirling-and-contraction} and the convexity of upper bounds obtained from degradable extensions then gives the result.
\end{proof}
\begin{figure}[h!]
  \centering
    \includegraphics[width=1.0\textwidth]{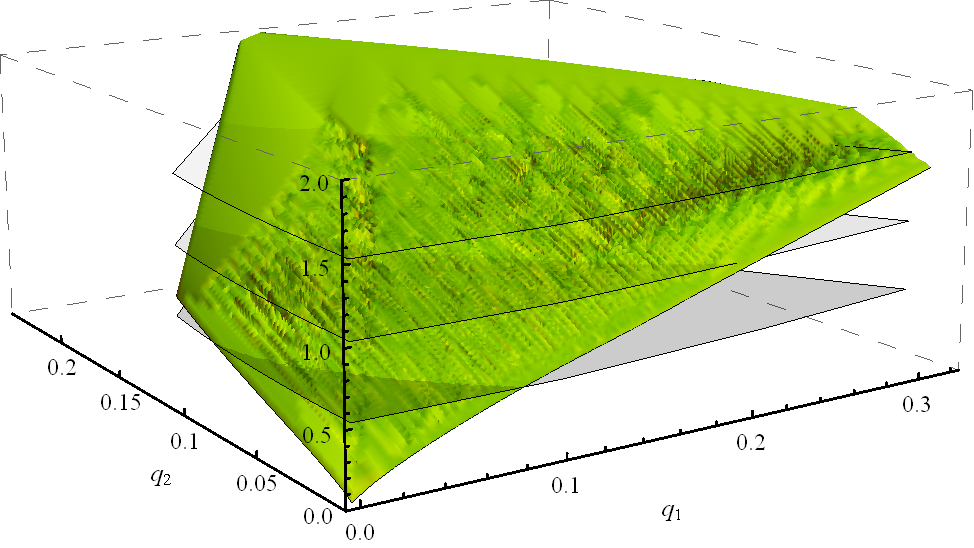}
      \caption{The concave roof of the depicted dimpled surface is our lower bound of $(2- Q(\cN))$ where $\cN$ is a $(q_1,q_2)$-channel (see Theorem \ref{thm:two-qubit-pauli}).  
      }
      \label{fig:locally-symmetric-channels}
\end{figure}

\subsection{Shifted Qubit Depolarizing Channels}
Various non-unital and non-degradable channels have interesting information theoretic properties \cite{RSW02, Fuk05, GLR05, CGMR08}, and it is natural to obtain upper bounds on their quantum capacities as well. We demonstrate that it is possible to obtain non-trivial upper bounds on the quantum capacity of a special non-unital and non-degradable qubit channel -- the shifted qubit depolarizing channel \cite{Fuk05,GLR05}.

\begin{figure}[h!]
  \centering
    \includegraphics[width=1.0\textwidth]{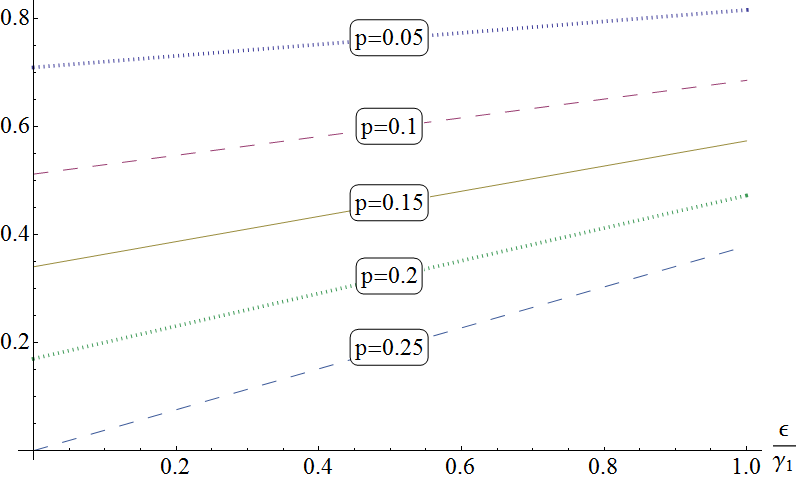}
      \caption{Upper bounds on $Q(\cD_{p,2,\eps \tZ})$ are depicted for different values of depolarizing probabilities $p$. Here $\gamma_1$ is a function of $p$ as defined in Theorem \ref{thm:shifteddepol2}.
      }
      \label{fig:shifted-depolarizing}
\end{figure}

The shifted depolarizing channel \cite{Fuk05,GLR05} of dimension $d$ is defined by
\begin{align}
\cD_{p,d,{\tens A}}(\rho) := \cD_{p,d}(\rho) + \tens A
\end{align}
where $\tens A$ is a $d$-dimensional Hermitian traceless matrix such that $\cD_{p,d,{\tens A}}$ is a completely positive map and hence still a quantum channel. Here, the operator $\tens A$ quantifies the amount by which the depolarizing channel $\cD_{p,d}$ is shifted. 
In the following theorem, we provide explicit upper bounds for the quantum capacity of the shifted qubit depolarizing channel (see also Figure \ref{fig:shifted-depolarizing}). To prove the theorem, we have to perform a specialized twirl on the qubit amplitude damping channel; this twirl is not the Pauli-twirl.
\begin{theorem} \label{thm:shifteddepol2}
For $ 0 < p  \le \frac{1}{4}$, let 
\[\gamma_1 = \sqrt{16-9p} + \frac{9p-16}{4}\]
 and 
 \[ \gamma_2 = 4\sqrt{1-p}(1-\sqrt{1-p}).\]
  Also
let 
\[g_1(p) =  H_2\bigl(\frac{1-\gamma_2}{2}\bigr) - H_2\bigl(\frac{\gamma_2}{2}\bigr),\]
 $g_2(p) = 1-H_2(p)$, and $g_3(p) = 1 - 4p$.
Then for all $\eps$ in the interval $[0,\gamma_1]$, we have
\begin{align}
Q(\cD_{p,2,\eps \tZ}) \le
 \eps \gamma_1^{-1} 
 \max_{q\in[0,1]} \Bigl\{
 I_{\rm coh}\bigl(\cA_{\gamma_1,2}, \diag(1-q,q) \bigr)
 \Bigr\}  
  + (1-\eps \gamma_1^{-1}) \conv(g_1,g_2,g_3; p, [0,\tfrac{1}{4}]) \notag.
\end{align}
\end{theorem}
\begin{proof}
Let $\mathcal U$ be the set of unitaries 
$\{  \mathbb 1,  
\frac{\tX + \tZ}{\sqrt{2}} ,  
\frac{\tY + \tZ}{\sqrt{2}}  
 \}$. 
Then the $\cU$-twirl of $\cA_{\gamma_1,2}$ is a shifted depolarizing channel, in the sense that 
\begin{align}
(\cA_{\gamma_1, 2})_{\ltimes \cU \rtimes} (\mathbb 1) &= \mathbb 1 + \gamma_1 \tens Z \notag\\
(\cA_{\gamma_1, 2})_{\ltimes \cU \rtimes} (\tP) &= \frac{2\sqrt{1-\gamma_1} +(1-\gamma_1)}{3} \tP \notag
\end{align}
for all non-trivial Paulis $\tP \in \{\tX,\tY,\tZ\}$. Thus $(\cA_{\gamma_1,2})_{\ltimes \cU \rtimes} = \cD_{p,2, \gamma_1 \tZ}$ where
\[
p  = \frac{4}{3}\left(1-\frac{2\sqrt{1-\gamma_1} +(1-\gamma_1)}{3}\right).
\]
Solving for non-negative $\gamma_1$ in terms of $p \in (0,\frac{1}{4}]$, we get
$
\gamma_1 = \sqrt{16-9p} + \frac{9p-16}{4}.
$
Hence
$
\cD_{p,2,\epsilon \tens Z} 
      = \eps \gamma_1^{-1} (\cA_{\gamma_1, 2})_{\ltimes \mathcal U \rtimes } + (1- \eps \gamma_1^{-1}) \cD_{p,2}.
$
Now $Q((\cA_{\gamma_1, 2})_{\ltimes \mathcal U \rtimes }) \le I_{\rm coh}(\cA_{\gamma_1, 2})$. By the method of degradable extension, $Q(\cD_{p,2}) \le \conv(g_1,g_2,g_3; p, [0,\tfrac{1}{4}])$ \cite{SS08}, and the result follows from the convexity of the upper bounds.
\end{proof}
\section{Concluding Remarks} \label{sec:degradable-conclude}
In this paper, we have generalized Smith and Smolin's result (Lemma 8 of \cite{SS08}) to our Theorem \ref{thm:twirling-and-contraction}, thereby upper bounding the quantum capacity of $\cV$-twirled degradable channels by their coherent information maximized on $\cV$-contracted input states. In essence, our main result elucidates a relationship between channel twirling, channel covariance and channel contraction. Additionally, we used our result to provide new upper bounds for the quantum capacity of several families of quantum channels using generalizations of the qubit amplitude damping channels as our ingredients. 

\section{Acknowledgements}
I thank Debbie Leung, Easwar Magesan, Graeme Smith, Mary Beth Ruskai, Mark Wilde, and the anonymous referees for their helpful comments. I acknowledge support from the Ministry of Education, Singapore.

\section{Appendix} \label{sec:degradable-appendix}
In this section, we explain some technical details in greater detail.
\subsection{Matrix Elements in the Pauli-basis}
Observe that
\begin{align}
4|0\>\<3| &= \tens{X} \otimes \tens{X} - \tens{Y} \otimes \tens{Y} + i (\tens{X} \otimes \tens{Y} + \tens{Y} \otimes \tens{X}) \label{03}\\
4|1\>\<2| &= \tens{X} \otimes \tens{X} + \tens{Y} \otimes \tens{Y} + i (-\tens{X} \otimes \tens{Y} + \tens{Y} \otimes \tens{X}) \label{12}\\
4|0\>\<2| &= \tens{X} \otimes \tens{\mathbb 1} + \tens{X} \otimes \tens{Z} + i (\tens{Y} \otimes \tens{\mathbb 1} + \tens{Y} \otimes \tens{Z}) \label{02} \\
4|1\>\<3| &= \tens{X} \otimes \tens{\mathbb 1} - \tens{X} \otimes \tens{Z} + i (\tens{Y} \otimes \tens{\mathbb 1} - \tens{Y} \otimes \tens{Z}) \label{13}\\
4|0\>\<1| &= \tens{\mathbb 1} \otimes \tens{X} + \tens{Z} \otimes \tens{X} + i (\tens{\mathbb 1} \otimes \tens{Y} + \tens{Z} \otimes \tens{Y}) \label{01}\\
4|2\>\<3| &= \tens{\mathbb 1} \otimes \tens{X} - \tens{Z} \otimes \tens{X} + i (\tens{\mathbb 1} \otimes \tens{Y} - \tens{Z} \otimes \tens{Y}).\label{23}
\end{align}
Also
\begin{align}
4 |0\>\<0| &= \tens{\mathbb 1} \otimes \tens{\mathbb 1} + \tens{\mathbb 1} \otimes \tens{Z} + \tens{Z} \otimes \tens{\mathbb 1} + \tens{Z} \otimes \tens{Z} \label{00}\\
4 |1\>\<1| &= \tens{\mathbb 1} \otimes \tens{\mathbb 1} - \tens{\mathbb 1} \otimes \tens{Z} + \tens{Z} \otimes \tens{\mathbb 1} - \tens{Z} \otimes \tens{Z} \label{11}\\
4 |2\>\<2| &= \tens{\mathbb 1} \otimes \tens{\mathbb 1} + \tens{\mathbb 1} \otimes \tens{Z} - \tens{Z} \otimes \tens{\mathbb 1} - \tens{Z} \otimes \tens{Z} \label{22}\\
4 |3\>\<3| &= \tens{\mathbb 1} \otimes \tens{\mathbb 1} - \tens{\mathbb 1} \otimes \tens{Z} - \tens{Z} \otimes \tens{\mathbb 1} + \tens{Z} \otimes \tens{Z} .\label{33}
\end{align}
We can also rewrite the above matrices in the following form.
\begin{align}
|0\>\<3| &= (|0\>\<0| )  (\tens{X} \otimes \tens{X}) \label{n03}\\
|0\>\<2| &= (|0\>\<0| )  (\tens{X} \otimes \tens{Z}) \label{n02}\\
|1\>\<3| &= (-|1\>\<1| ) (\tens{X} \otimes \tens{Z}) \label{n13}\\
|0\>\<1| &= (|0\>\<0| )  (\tens{Z} \otimes \tens{X}) \label{n01}\\
|2\>\<3| &= (-|2\>\<2| ) (\tens{Z} \otimes \tens{X}) \label{n23}
\end{align}
\subsection{Proof of Proposition \ref{prop:twirled-extension-C}}
\label{sec:proof-prop-twirled-ext}
 Let $\fK_{\cN}$ denote the Kraus set of the channel $\cN$. Using the canonical definition of the complementary channel of $\cN$ from its canonical isometric extension, we have for all $\tV \in \cV$,
 \begin{align}
 \cN^C(\tV \rho \tV^\dagger)
     = 
  \tr_{\cH_B} \left (
        \Bigl (
            \sum_{  \tA,  \tA' \in \fK_\cN   }
                \tA \tV \rho \tV^\dagger \tA'^\dagger
        \Bigr )_{\cH_B}
       \otimes |   \tA   \> \<   \tA'  |
  \right ).
 \end{align}
Similarly, the canonical complementary channel of $\widetilde \cN$ is 
 \begin{align}
 \widetilde \cN^C(\rho)  &=  
 \tr_{\cH_B \otimes \cH_C} \left(
   \frac{1}{|\cV|} 
    \sum_{\substack{V,V' \in \cV \\
                               A,A' \in \fK_\cN}}
      \Bigl (
           \tV^\dagger \tA \tV 
                   \rho 
           \tV^\dagger \tA'^\dagger \tV'
        \Bigr )_{\cH_B}
       \otimes  \Bigl( |   \tV   \> \<   \tV'  | \Bigr)_{\cH_C}
        \otimes |   \tA   \> \<   \tA'  |
         \otimes |   \tV   \> \<   \tV'  |
\right) \notag\\
&= 
   \frac{1}{|\cV|} 
   \sum_{V \in \cV} \tr_{\cH_B } \left(
   \sum_{A,A' \in \fK_\cN}
      \Bigl (
           \tV^\dagger \tA \tV 
                   \rho 
           \tV^\dagger \tA'^\dagger \tV
        \Bigr )_{\cH_B}
        \otimes |   \tA   \> \<   \tA'  | \right)
         \otimes |   \tV   \> \<   \tV  | \notag\\
&= 
   \frac{1}{|\cV|} 
   \sum_{V \in \cV} 
   \cN^C( \tV 
                   \rho 
           \tV^\dagger )
         \otimes |   \tV   \> \<   \tV  | \notag
 \end{align}
 where we have used the unitary invariance of the partial trace.
\hfill $\qed$
\subsection{Proof of Lemma \ref{lem:Phixyzdegradable} } \label{sec:proof-degradable}
When $(x,y,z) \in \fF_{x,y,z}$, the map $\cA_{x,y,z}$ is a quantum channel and $g,h,k \in [0,1]$.
 Hence $\cA_{g,h,k}$ is also a quantum channel. 
Also note that $\cA_{x,y,z}^C = \cA_{1-x,y,1-2y-z}$.

We now proceed to show that $\cA_{g,h,k} \circ \cA_{x,y,z}  = \cA_{x,y,z}^C$ which will imply that $\cA_{x,y,z}$ is a degradable channel. We denote the Kraus operators of $\cA_{x,y,z}$, $\cA_{x,y,z}^C$ and $\cA_{g,h,k}$ by $\tA_i$, $\tR_i$ and $\tG_i$ respectively, where $\tA_i$ is given by (\ref{channel-definition}), 
\begin{align*}
\tens{R}_0 &= |0\>\<0| +\sqrt{x} |1\>\<1| + \sqrt{x} |2\>\<2| +\sqrt{z}|3\>\<3| \notag\\
\tens{R}_1 &= \sqrt{1-x} |0\>\<1| + \sqrt{y} |2\>\<3|\notag\\
\tens{R}_2 &= \sqrt{1-x} |0\>\<2| + \sqrt{y} |1\>\<3| \notag\\
\tens{R}_3 &= \sqrt{1-2y -z} |0\>\<3|,
\end{align*}
and
\begin{align*}
\tens{G}_0 &= |0\>\<0| + \sqrt{1-g} ( |1\>\<1| +|2\>\<2|) + \sqrt{1-2h-k}|3\>\<3| \\
\tens{G}_1 &= \sqrt{g} |0\>\<1| + \sqrt{h} |2\>\<3|\\
\tens{G}_2 &= \sqrt{g} |0\>\<2| + \sqrt{h} |1\>\<3|\\
\tens{G}_3 &= \sqrt{k} |0\>\<3|.
\end{align*}
   By the Kraus representation,
$
\displaystyle
\mathcal \cA_{g,h,k} (\cA_{x,y,z}(\rho )) = \sum_{k,\ell \in \{0,1,2,3\} } \tens{G}_k \tA_\ell \rho \tens{A}_\ell^\dagger \tens{G}_k^\dagger.
$
In this representation, the composite quantum channel $\cA_{g,h,k} \circ \cA_{x,y,z}$ has sixteen Kraus operators
$\tens{G}_k \tens{A}_\ell$
for $k , \ell \in \bbZ_4$. Now we evaluate $\tens{G}_k \tens{A}_\ell$ explicitly.
\begin{align*}
\tens{G}_1 \tens{A}_3 = \tens{G}_1 \tens{A}_1 = 0 ,  \tens{G}_1 \tens{A}_2 = \sqrt{\frac{1-2x}{1-x} y} |0\>\<3| \\
\tens{G}_2 \tens{A}_3 = \tens{G}_2 \tens{A}_2 = 0,   \tens{G}_2 \tens{A}_1 = \sqrt{\frac{1-2x}{1-x} y } |0\>\<3|\\
\tens{G}_3 \tens{A}_3 = \tens{G}_3 \tens{A}_2 =      \tens{G}_3 \tens{A}_1 = 0.
\end{align*}
Also we have
\begin{align*}
\tens{G}_1 \tA_0 = \sqrt{1-2x} |0\>\<1| + \sqrt{\frac{1-2x}{1-x} y} |2\>\<3| \\
\tens{G}_2 \tA_0 = \sqrt{1-2x} |0\>\<2| + \sqrt{\frac{1-2x}{1-x} y} |1\>\<3| \\
\tens{G}_3 \tA_0 = \sqrt{\frac{1-x-2y(2-3x)}{1-x}  } |0\>\<3|.
\end{align*}
Moreover
\begin{align*}
\tens{G}_0 \tA_1 = \sqrt{x}|0\>\<1| + \sqrt{\frac{x y}{1-x}} |2\>\<3|\\
\tens{G}_0 \tA_2 = \sqrt{x}|0\>\<2| + \sqrt{\frac{x y}{1-x}} |1\>\<3|\\
\tens{G}_0 \tA_3 = \sqrt{z}|0\>\<3|.
\end{align*}
Observe then that $\tens{G}_0 \tA_1 = \sqrt{\frac{x}{1-2x}} \tens{G}_1 \tA_0$ 
and $\tens{G}_0 \tA_2 = \sqrt{\frac{x}{1-2x}} \tens{G}_2 \tA_0$. Thus applying 
the Kraus operators $\tens{G}_i \tA_0$ and $\tens{G}_0 \tA_i$ is equivalent to 
applying the Kraus operator $\tens{R}_i$ for $i \in \{1,2\}$. Similarly, applying
 the Kraus operators $\tens{G}_1 \tA_2, \tens{G}_2 \tA_1$ and $\tens{G}_3 \tA_0$ 
 is equivalent to applying the Kraus operator $\tens{R}_3$. Moreover,
  since $1-g = \frac{x}{1-x}$ and $(1-2h-k)(1-2y-z)=z$, we have that $\tens{G}_0 \tA_0 = \tens{R_{0}}$.
  Hence $\cA_{g,h,k}\circ \cA_{x,y,z}  = \cA_{x,y,z}^C$.
   \hfill  $\qed $

\subsection{Twirling of Channels}
To obtain locally symmetric Pauli channels, we introduce the notion of localized Clifford twirling. Instead of twirling our channel over the entire Clifford group over all the qubits \cite{DLT02}, we can twirl the channel with respect to the Clifford group for individual qubits independently. The material below is an explicit discussion on the notion of localized Clifford twirling.

Now define the set of non-trivial Pauli matrices to be $\mathcal P^*_1 := \{\tens{X}, \tens{Y}, \tens{Z}\}$.  We study a set of automorphisms on the non-trivial Pauli matrices. To define this set of automorphisms, we first define a Hermitian and traceless qubit operator
\begin{align*}
\tens{H}_{\tau_1,\tau_2} := \frac{\tau_1+\tau_2}{\sqrt{2}}
\end{align*}
for all non-trivial Pauli matrices $\tau_1$ and $\tau_2$, which is just the Hadamard matrix in an arbitary Pauli basis.
For all non-trivial Pauli matrices $\tW$, conjugation of $\tens{W}$ with $\tens{H}_{\tau_1,\tau_2}$ gives the following.
\begin{align*}
\tens{H}_{\tau_1,\tau_2} \tens{W} \tens{H}_{\tau_1,\tau_2} =
\left \{
\begin{array}{ll}
\tau_1 &, \quad  \tens{W} = \tau_2 \\
\tau_2 &, \quad \tens{W} = \tau_1 \\
-\tens{W} &, \quad \tens{W} \notin \{\tau_1 , \tau_2 \} \\
\end{array}
\right.
\end{align*}
Hence the automorphism associated with the generalized Hadamards $\tens{H}_{\tau_1, \tau_2}$ on the set of non-trivial Pauli matrices swaps $\tau_1$ and $\tau_2$. The size of the set of all automorphisms on the set of non-trivial Pauli matrices is the size of the symmetric group of order 3, which is 6. Hence we consider the set
\begin{align}
\mathcal B := \{ \tens{\mathbb 1}, \tens{H_{\tens{X},\tens{Y}}}, \tens{H_{\tens{X},\tens{Z}}}, \tens{H_{\tens{Y}, \tens{Z}}}, \tens{H_{\tens{X}, \tens{Z}}} \tens{H_{\tens{X}, \tens{Y}}}, \tens{H_{\tens{X}, \tens{Y}}} \tens{H_{\tens{X}, \tens{Z}}} \} \label{eq:cBset}
\end{align}
with six qubit operators, each operator corresponding to a distinct automorphism of the set of non-trivial Pauli matrices.
For all  $\tens{P}, \tens{V} \in \mathcal P_1$, observe that
\begin{align}
&\frac{1}{6}\sum_{\tens{B} \in \mathcal B} (\tens{B^\dagger PB} ) \tens{V} (\tens{B^\dagger P B  }) = \left\{
\begin{array}{ll}
\frac{1}{3}\sum_{\tens{P'} \in \mathcal P_1^*} \tens{P'} \tens{V} \tens{P'} & , \quad \tens{P} \in \mathcal P^*_1  \\
\tens{V} & , \quad \tens{P} = \tens{\mathbb 1} \\
\end{array}\right.
\label{eq:twirl-id1}.
\end{align}

\begin{proposition}\label{prop:twirling}
Let $\mathcal N$ be a two-qubit channel with Kraus set $\fK_\cN$ and 
\[{\displaystyle a_{\tens{P} \otimes \tens{P'}} = 
\frac{1}{16}   \sum_{\tens K \in \fK_\cN } 
 \Bigl|\tr( (\tP \otimes \tP' )\tens{K})  \Bigr|^2      }.
\]
Then $((\cN_{\ltimes \cP_2 \rtimes })_{\ltimes \cB \otimes \bbone \rtimes })_{\ltimes \bbone\otimes \cB \rtimes}$ is a two-qubit Pauli channel with Kraus operators 
$\sqrt{a_{\bbone \otimes \bbone}}\bbone \otimes \bbone$, 
$
\Bigl( {\displaystyle \sum_{\tR \in \cP^*_1}\frac{1}{3}} a_{\tR \otimes \bbone }\Bigr)^{\frac{1}{2}}
 {\tR \otimes \bbone }$, 
$\displaystyle 
\Bigl( \sum_{\tR \in \cP^*_1}\frac{1}{3} a_{\bbone \otimes \tR} \Bigr)^{\frac{1}{2}} \bbone \otimes \tR $, and 
$\displaystyle 
\Bigl (\sum_{\tR,\tR' \in \cP^*_1}\frac{1}{9} a_{\tR \otimes \tR' } \Bigr)^{\frac{1}{2}}
\tR \otimes \tR'
$ respectively where $\tR, \tR' \in \cP_1$. Moreover if $\cN = \cA_{x,y,z}$, then $((\cN_{\ltimes \cP_2 \rtimes })_{\ltimes \cB \otimes \bbone \rtimes })_{\ltimes \bbone\otimes \cB \rtimes}$ is a $(q_1,q_2)$-channel with $q_1$ and $q_2$ given by (\ref{eq:q1q2-xyz}).
\end{proposition}
\begin{proof}
Let $\tens{V}$ and $\tens{W}$ be single qubit Pauli matrices. Then using (\ref{eq:twirl-id1}) we get
\begin{align}
\mathcal N_{\wr \cB \otimes \mathbb 1 \wr}(\tens{V} \otimes \tens{W})
=&
\frac{1}{6}\sum_{\tens{B} \in \mathcal B}  \sum_{\tens{P, P'} \in \mathcal P_1}
\tens{B^\dagger  P B V B^\dagger P B}  \otimes \tens{P' W P' } a_{\tens{P} \otimes \tens{P'}} \notag\\
=&
\frac{1}{6}\sum_{\tens{P, P'} \in \mathcal P_1} 
\Bigl(\sum_{\tens{B} \in \mathcal B}
(\tens{B^\dagger  P B} ) \tens{V} ( \tens{B^\dagger P B} ) \Bigr)
  \otimes \tens{P' W P'} a_{\tens{P} \otimes \tens{P'}} \notag\\
=&
\sum_{\tens{P'} \in \mathcal P_1} \tens{V}   \otimes \tens{P'} \tens{W} \tens{P'} a_{\tens{\mathbb 1} \otimes \tens{P'}}
 + \frac{1}{3}\sum_{\tens{P} \in \mathcal P_1^*} \Bigl(\sum_{\tens{R} \in \mathcal P^*_1}
\tens{R V R} \Bigr)  \otimes \sum_{\tens{P'} \in \mathcal P_1} \tens{P' W P'} a_{\tens{P} \otimes \tens{P'}} \notag.
\end{align}
By rearranging the terms above, we get
\begin{align*}
\mathcal N_{\wr \cB \otimes \mathbb 1 \wr}(\tens{V} \otimes \tens{W})
=&
 \tV   \otimes \sum_{\tens{P'} \in \mathcal P_1} \tens{P' W P'} a_{\tens{\mathbb 1} \otimes \tens{P'}}
 +  \Bigl(\sum_{\tens{R} \in \mathcal P^*_1}
\tens{R V R} \Bigr)  \otimes \sum_{\tens{P'} \in \mathcal P_1} \tens{P' W P'}  \sum_{\tens{P} \in \mathcal P_1^*} \frac{a_{\tens{P} \otimes \tens{P'}}}{3}.
\end{align*}
Similarly,
\begin{align*}
(\mathcal N_{\wr \cB \otimes \mathbb 1 \wr})_{\wr \mathbb 1 \otimes \cB \wr}(\tens{V} \otimes \tens{W})
=&
\tV  \otimes \frac{1}{6} \sum_{ \tens{B} \in \mathcal B }  \sum_{\tens{P'} \in \mathcal P_1}
(\tens{B^\dagger P'B})\tens{ W}
(\tens{B^\dagger P'B}) a_{\tens{\mathbb 1} \otimes \tens{P'}} \notag\\
& +  \Bigl(\sum_{\tens{R} \in \mathcal P^*_1}
\tens{R V R} \Bigr)  \otimes \frac{1}{6} \sum_{ \tens{B} \in \mathcal B }
\sum_{\tens{P'} \in \mathcal P_1} (\tens{B^\dagger P'B}) \tens{W }(\tens{B^\dagger P'B})
 \Bigl( \sum_{\tens{P} \in \mathcal P_1^*} \frac{a_{\tens{P} \otimes \tens{P'}}}{3}\Bigr)\notag\\
=&
a_{\tens{\mathbb 1} \otimes \tens{\mathbb 1}}\tens{ V}   \otimes \tens{W} + \tens{ V } \otimes 
\Bigl(\sum_{ \tens{R'} \in \mathcal P_1^* }  \tens{R' W R'} \Bigr)  
\Bigl( \sum_{\tens{P'} \in \mathcal P_1^*} \frac{a_{\tens{\mathbb 1} \otimes \tens{P'}}}{3} \Bigr) 
\notag\\
& +  
\Bigl(\sum_{\tens{R} \in \mathcal P^*_1} \tens{R V R} \Bigr) 
 \otimes \tens{W}
  \Bigl( \sum_{\tens{P} \in \mathcal P_1^*} \frac{a_{ \tens{P} \otimes \tens{\mathbb 1}}}{3} \Bigr)
   \notag\\
& + \Bigl(\sum_{\tens{R} \in \mathcal P^*_1} \tens{R V R} \Bigr)  \otimes \Bigl(\sum_{\tens{R'} \in \mathcal P^*_1} \tens{R' W R'} \Bigr)
 \sum_{\tens{P,P'} \in \mathcal P_1^*} \frac{a_{\tens{P} \otimes \tens{P'}}}{9}. 
\end{align*}
This completes the first part of the proof.

Now the Pauli-twirl of $\cA_{x,y,z}$ has the Kraus operators
\begin{align}
    &\left( \frac{1+2\sqrt{1-x} + \sqrt{1-2y} }{4} \right) \mathbb 1 \otimes \mathbb 1,\notag\\
    & \left( \frac{1-\sqrt{1-2y} }{4}\right) \tens P , & \tens P \in \{ \mathbb 1 \otimes \tens Z, \quad \tens Z \otimes \mathbb 1\} \notag\\
    &\left| \frac{1-2\sqrt{1-x} + \sqrt{1-2y}}{4}\right| \tens Z \otimes \tens Z \notag\\
    & \left| \frac{\sqrt{x} + \sqrt{y}}{4}\right| \tens P ,
     & \tens P \in \{ \mathbb 1 \otimes \tens X,\quad  \mathbb 1 \otimes \tens Y, \quad \tens X \otimes \mathbb 1, \   \quad \tens Y \otimes \mathbb 1 \} \notag\\
    & \left|\frac{\sqrt{x}-\sqrt{y}}{4}\right| \tens P,
    & \tens P \in \{ \tens Z \otimes \tens X, \quad \tens Z \otimes \tens Y, \quad \tens X \otimes \tens Z  , \quad \tens Y \otimes \tens Z  \} \notag \\
    &\frac{\sqrt{z}}{2} \tens P,
    & \tens P \in \{ \tens{X} \otimes \tens{X} , \quad \tens{X} \otimes \tens{Y} , \quad \tens{Y} \otimes \tens{X}, \quad \tens{Y} \otimes \tens{Y} \} \notag
\end{align}
and hence combining this with the first result of our proposition, the second result of our proposition follows. \hfill 
\end{proof}

\bibliography{mybib}{}
\addcontentsline{toc}{paper}{Bibliography}
\bibliographystyle{ieeetr}

\end{document}